\documentclass[aps,prd,onecolumn,groupedaddress,showpacs,nofootinbib,amssymb
]{revtex4}
\usepackage[dvips]{graphicx}
\usepackage{amssymb}
\usepackage{amsmath}
\usepackage{graphicx}
\usepackage{amsfonts}
\usepackage{bm}

\begin{document}

\title{Big-Bounce with Finite-time Singularity: The $F(R)$ Gravity Description}
\author{
S.~D.~Odintsov,$^{1,2}$\,\thanks{odintsov@ieec.uab.es}
V.~K.~Oikonomou,$^{3,4}$\,\thanks{v.k.oikonomou1979@gmail.com}}
\affiliation{ $^{1)}$Institut de Ciencies de lEspai (IEEC-CSIC),
Campus UAB, Carrer de Can Magrans, s/n\\
08193 Cerdanyola del Valles, Barcelona, Spain\\
$^{2)}$ ICREA, Passeig LluA­s Companys, 23,
08010 Barcelona, Spain\\
$^{3)}$ Tomsk State Pedagogical University, 634061 Tomsk, Russia\\
$^{4)}$ Laboratory for Theoretical Cosmology, Tomsk State University of Control Systems
and Radioelectronics (TUSUR), 634050 Tomsk, Russia\\
}

\begin{abstract}
An alternative to the Big Bang cosmologies is obtained by the
Big Bounce cosmologies. In this paper, we study a bounce cosmology with a
Type IV singularity occurring at the bouncing point, in the context of $F(R)$ modified gravity.
We investigate the evolution of the Hubble radius and we examine the issue of primordial cosmological
perturbations in detail. As we demonstrate, for the singular bounce, the primordial perturbations originating from
the cosmological era near the bounce, do not produce a scale invariant spectrum and also the short wavelength modes,
 after these exit the horizon, do not freeze, but grow linearly with time. After presenting the cosmological perturbations study,
 we discuss the viability of the singular bounce model, and our results indicate that the singular bounce must be combined with
 another cosmological scenario, or should be modified appropriately, in order that it leads to a viable cosmology. The study
  of the slow-roll parameters leads to the same result, indicating the singular bounce theory is unstable at the singularity point,
  for certain values of the parameters. We also conformally transform the Jordan frame singular bounce, and as we demonstrate, the
   Einstein frame metric leads to a Big Rip singularity. Therefore, the Type IV singularity in the Jordan frame, becomes a Big Rip
   singularity in the Einstein frame. Finally, we briefly study a generalized singular cosmological model, which contains two Type
   IV singularities, with quite appealing features.
\end{abstract}


\maketitle



\def\pp{{\, \mid \hskip -1.5mm =}}
\def\cL{\mathcal{L}}
\def\be{\begin{equation}}
\def\ee{\end{equation}}
\def\bea{\begin{eqnarray}}
\def\eea{\end{eqnarray}}
\def\tr{\mathrm{tr}\, }
\def\nn{\nonumber \\}
\def\e{\mathrm{e}}

\section{Introduction}

One of the very interesting questions posed in modern theoretical cosmology is whether the Universe was created in an initial singular state, or whether the Universe oscillates in a bouncing-like way, so the dilemma is, did the Universe started with a Big Bang or with a Big Bounce? Put in other words, does the Universe evolves according to the standard inflationary paradigm, for which an initial singularity exists, or the Universe is described by a cosmological bounce, in which case the Universe has no initial singularity. This question is one of the most interesting questions in modern theoretical cosmology, and in order to be answered correctly, a lot of different theoretical disciplines has to be used, and also supplemented by observational data. The successes of the standard inflationary scenario are quite many, for example the horizon and flatness problems are resolved, see for example \cite{inflation,pert} for some review articles and important papers on this vast research topic. However, the recent observations \cite{planck}, do not necessarily prove that inflation indeed took place. Particularly, the observations indicate that the power spectrum of primordial curvature fluctuations is nearly scale invariant, but this is not an indication of an inflationary era. On the antipode of the inflationary paradigm stands the big bounce cosmological theory, which supports that the Universe undergoes in two eras of acceleration, the contracting era and the expansion era, and in between these eras, the Universe reaches a minimum size, at which it bounces off. In the literature, the bounce cosmology paradigm has been extensively studied during the last twenty years, for an incomplete list, see for example \cite{bounce1,bounce2,bounce3,bounce4,bounce5,quintombounce,ekpyr1,superbounce2,matterbounce,mukhanov1aa}. In the context of bouncing cosmology, it is also possible to have a nearly scale invariant power spectrum of primordial curvature perturbations, as is the case for the matter bounce scenario \cite{matterbounce,mukhanov1aa}. But the most appealing feature of the big bounce paradigm is that the initial singularity problem, from which the Big Bang scenario suffers, is consistently alleviated.

In principle, the singularities in physical theories and in general relativity, can be viewed as an ``embarrassing'' feature for the physical description, since the physical description actually fails to describe consistently the physical phenomenon at the singularity point. In some sense it reveals that the physical description at the singularities fails, so a new more fundamental theory is needed, in the context of which, the singularity will not appear. One characteristic example of this is the singular potential for a point charge, like the electron, in which case the singularity is resolved in the quantum electrodynamics description, since the electron self-interaction consistently solves the problem.

In general relativity though, there exist various types of singularities, with most commonly studied being the singularities of black holes. The singularities occurring in gravitational collapse, like the singularities in black holes, are spacelike singularities, and received much attention and publicity, mainly because of the fascinating and appealing status of black holes. In cosmology though there are also finite time singularities, which are timelike singularities, and are not so much publicly known, except for the initial singularity in the Big Bang scenario. All types of singularities are described very consistently by the Hawking-Penrose theorems \cite{hawkingpenrose}. The spacelike singularities commonly occur during the gravitational collapse of massive objects \cite{Virbhadra:2002ju}, while timelike singularities have a more ``cosmic'' effect, since these lead to global phenomena which affect all physical quantities which can be defined on a three dimensional spacelike hypersurface. Usually the spacelike singularities are crushing types singularities and for which the curvature scalars strongly diverge. However, in cosmology there are various types of finite-time singularities, which were firstly classified and systematically studied in \cite{Nojiri:2005sx}. The most severe from a phenomenological point of view is the Big Rip singularity \cite{ref5}, but there are three more singularities, the Type II, III and Type IV singularities. All except from the last one, lead to singular physical quantities on the three dimensional spacelike hypersurface that is defined by the singularity point. However, the last one, the Type IV does not lead to catastrophic events with regards to the observable quantities. For recent and earlier studies on this type of singularity, see for example \cite{Nojiri:2005sx,Barrow:2015ora,noo1,noo2,noo3,noo4,noo5,noo6}. Milder singularities of this type were studied initially in Ref. \cite{barrownew}, where it was shown that, contrary to what expected at that time, closed universes obeying $\rho + 3p>0$ and $\rho>0$, need not to recollapse, due to the fact that the Universe might experience a pressure singularity before an expansion maximum of the Universe is achieved. Later the milder singularities types, were developed in \cite{Barrow:2004xh} and \cite{barrow}, where they were called sudden singularities. The Type IV singularity has no effect on the observable quantities, but however it affects severely the dynamical evolution of the cosmological system, as was demonstrated in \cite{noo4,noo5,noo6}. One big difference between the crushing type singularities and the Type IV singularity is that for the former ones, geodesics incompleteness occurs at the singularity point, a feature which is of course absent for the Type IV singularity. Therefore, the Universe will smoothly pass through the singular point and only it's dynamics will be affected, but not the physical quantities that characterize it. Specifically, when we refer to dynamics, we mean the slow-roll parameters which determine the dynamical evolution of the cosmological system \cite{barrowslowroll}. As was demonstrated in \cite{noo3,noo4,noo5}, for a Type IV singularity it is possible that the slow-roll parameters diverge and hence the system becomes dynamically unstable, since the slow-roll expansion breaks down at the singularity, and possibly at a higher order of the slow-roll expansion. This instability may indicate that graceful exit from inflation may be triggered in some cases, see for example \cite{noo4,noo5}. In addition, for a non-singular bounce, it was demonstrated in \cite{sergeiharo}, that the Big Rip singularity is avoided, and also the Type II and III singularities are also avoided, see \cite{sergeiharo} for details.

In this paper we shall be interested in the simplest form of a
cosmological bounce model which will incorporate a Type IV
singularity. Particularly, the Type IV singularity will be chosen to
occur at exactly the bouncing point. In Ref. \cite{noo5} we
investigated how this singular bounce can be realized in the context
of vacuum $F(R)$ modified gravity (for review, see \cite{reviews1}),
and we use the results of that work to further analyze the singular
bounce in the Jordan frame. Our first task in this work is to study
the power spectrum of primordial curvature perturbations. Of course
before calculating the power spectrum and the corresponding spectral
index, we investigate the specifics of the singular bounce
cosmological evolution, emphasizing on finding the era at which the
primordial perturbations are generated. As we evince, in the case at
hand, the cosmological evolution is very different than the standard
inflationary paradigm, since the evolution of the Comoving Hubble
radius is totally different. Particularly, at the singularity point,
the Hubble horizon is infinite, and as time grows, it shrinks
eventually. The short wavelength modes, which are relevant for
present day's observations, are well inside the Hubble horizon at
the singularity point. These short wavelength modes eventually exit
the Hubble horizon at a later time, but still very near to the
singular bouncing point. The cosmological perturbations are
calculated, using standard techniques, and also by assuming that
these originate from quantum vacuum fluctuations. As we demonstrate,
the power spectrum of primordial curvature perturbations is not
scale invariant, and actually cannot be for any physical value of
the parameters used. This is one of the shortcomings of the singular
bounce scenario, since there are some features of the singular
bounce that make it not so appealing, at least if it is not combined
with other cosmological scenarios. Particularly, another not so
appealing feature is that when the short wavelength modes exit the
Hubble horizon, these actually do not freeze for the singular
bounce, but grow linearly with time. In addition, the fact that the
Hubble horizon continuously shrinks for the singular bounce,
indicates that the singular bounce can only be viable only if
combined with other cosmological scenarios, but by itself it yields
a not so appealing cosmological evolution. Of course we need to note
that our calculations were performed for cosmic times near the
bounce, and therefore our results are only approximations near the
bounce. Also, the calculation of the cosmological perturbations and
therefore of the corresponding power spectrum, was performed by
using a linear approximation for the perturbations. It is possible
that these approximations are not valid, and in such a case, the
singular bounce might describe an era before the era for which
cosmological perturbations with relevance to today's observations
are generated. For example, the singular bounce might describe the
quantum era, which is followed by another cosmological scenario, as
was done in Refs. \cite{lcdmcai} (see also \cite{lcdmsergei} for an
$F(R)$ gravity description). After the cosmological perturbations
study, we investigated how the slow-roll indices behave in the case
of a Type IV singularity. As we demonstrate, some of the slow-roll
indices blow-up at the singularity, and therefore this clearly
indicates that the cosmological system becomes dynamically unstable.
Notice that this does not mean that the observational indices become
singular, since the observational indices are expressed in terms of
the slow-roll indices, only when these are much smaller than one. In
addition, we investigate how the Jordan frame singular bounce
behaves when conformally transformed in the Einstein frame. As we
evince, the Type IV singularity of a Jordan frame $F(R)$ gravity
becomes a Big Rip singularity in the Einstein frame. We need to note
that the singular bounce we studied was the simplest case of a
cosmological evolution that can incorporate a Type IV singularity,
and as we demonstrated it does not suffice to produce a viable
cosmology. To this end, our final study for this paper involves a
model of cosmological evolution which has two Type IV singularities
occurring in two different time instances. We briefly discuss it's
qualitative features and we investigate how the equation of state
behaves for this model. As we demonstrate, the model can describe
both the early and late-time acceleration eras, but also the matter
domination era. The phenomenological consequences of the model are
particularly interesting since at early times it is approximated by
the Starobinsky $R^2$ inflation model \cite{starobinsky}, which is
compatible with the latest Planck data \cite{planck}. We need to
note that the present work is qualitatively but also quantitatively
different from our work \cite{noo4}, since form a qualitative point
of view, inflation is replaced by a bouncing cosmology and from a
quantitative point of view, the slow-roll indices in the present
work, do not suffice to determine the power spectrum of primordial
curvature perturbations, which have to be calculated explicitly by
calculating the comoving curvature perturbation from scratch.

The outline of the paper is as follows: In section II we present in brief all the essential information with regards to finite time singularities, and we also give a brief description of the simplest singular cosmological evolution, which is also a bouncing cosmology. In section III, we study in detail the cosmological perturbations of the singular bounce model in the context of $F(R)$ gravity. In addition, we discuss the flaws of the model and also suggest possible modifications that could render the model more appealing from a phenomenological point of view. Our study also involves a deep analysis of the Hubble horizon behavior for the singular bounce model. In section IV, we investigate how the singular bounce behaves in the Einstein frame, after conformally transforming the Jordan frame metric. In section V we study the behavior of the slow-roll indices in detail, by also calculating the Hubble slow-roll indices. We discuss the implications of the Type IV singularity on the slow-roll indices, and we stress the fact that the implications do not affect the observational indices, but only affect the dynamics. Finally, in section VI we present in brief a very appealing cosmological model, which has two Type IV singularities occurring at two different time instances. As we demonstrate, in this case the resulting Type IV singular theory may lead to a very promising and phenomenologically viable cosmological model. The conclusions along with a discussion on potential applications or modifications of our work, follow in the end of the paper.

The conventions we shall adopt in this paper are the following:  For simplicity we shall assume that the geometric background is a flat Friedmann-Robertson-Walker background, with it's line element being,
\be
\label{metricfrw} ds^2 = - dt^2 + a(t)^2 \sum_{i=1,2,3}
\left(dx^i\right)^2\, ,
\ee
with $a(t)$ denoting the scale factor. Moreover, we shall assume that the connection is a symmetric, torsion-less and metric compatible affine connection, the Levi-Civita connection. In addition, the Ricci scalar for the metric (\ref{metricfrw}) reads,
\begin{equation}
\label{ricciscal}
R=6(2H^2+\dot{H})\, ,
\end{equation}
where $H$ is the Hubble rate which is defined as $H=\dot{a}/a$.

\section{Essentials of Finite-time Singularities and Brief Description of the Singular Bounce}

\subsection{Finite-time Singularities and Geometric Conventions}

Before we describe the cosmological evolution of the singular big bounce model we shall study in this paper, it is worth recalling some essential features of the finite-time singularities and also describe in brief the singular bounce. For more details on the finite-time singularities, we refer the reader to Ref.~\cite{Nojiri:2005sx}.  The classification of finite-time singularities was firstly done in Ref. \cite{Nojiri:2005sx}, according to which we have four different types of finite-time singularities, which are the following:
\begin{itemize}
\item Type I (``Big Rip Singularity''): From the four types of finite-time singularities, it is the most ''severe'', with regards to it's phenomenological implications, since all the physical quantities that can be defined on a constant time, three dimensional spacelike hypersurface, diverge at the singularity instance. Particularly, the singularity is a timelike singularity which occurs as the cosmic time approaches a time instance $t_s$, and when $t\rightarrow t_s$, the scale factor $a(t)$ of the metric, the effective pressure
$p_\mathrm{eff}$ and the effective energy
density $\rho_{\mathrm{eff}}$ strongly diverge, that is, $a \to \infty$, $\left|p_\mathrm{eff}\right| \to
\infty$ and $\rho_\mathrm{eff} \to \infty$. For an important stream of research articles for this type of singularity, we refer to Refs.~\cite{ref5}.
\item Type II (``Sudden Singularity''): This type of singularity was firstly studied in \cite{Barrow:2004xh,barrow}, and in this case, the effective energy density $\rho_{\mathrm{eff}}$ and the scale factor $a(t)$ are both finite, and only the effective pressure diverges, that is  $a
\to a_s$, $\rho_{\mathrm{eff}}\to \rho_s$ but $\left|p_\mathrm{eff}\right| \to \infty$.
\item Type III: This is the second in order of the most severe singularities, and in this case, both the effective pressure and the effective energy density diverge, as $t\to t_s$, that is, $\left|p_\mathrm{eff}\right| \to \infty$ and $\rho_\mathrm{eff} \to \infty$, but the scale factor is finite, that is, $a \to a_s$.
\item Type IV: The most interesting due to it's phenomenological implications. In the case that a Type IV singularity occurs at some time instance, the Universe can smoothly pass through it, without the physical quantities becoming divergent. This singularity affects the dynamics of the cosmological evolution, since it may cause the slow-roll indices to diverge \cite{noo4,noo5,noo6}. For this singularity, the scale factor of the metric $a(t)$, the
energy density $\rho_\mathrm{eff}$, and finally, the pressure $\left|p_\mathrm{eff}\right|$ remain finite, that is,  $a \to a_s$,
$\rho_\mathrm{eff} \to \rho_s$ and
$\left|p_\mathrm{eff}\right| \to p_s$. Moreover, the Hubble rate and it's
first derivative are also finite, but only the higher derivatives of the Hubble rate diverge. For some recent studies on the implications of the Type IV singularity, we refer the reader to \cite{Barrow:2015ora,noo1,noo2,noo3,noo4,noo5,noo6}.
\end{itemize}

\subsection{A Brief Description of the Singular Bounce}

In this paper we shall assume that the Universe's evolution is described by a singular bounce, with Hubble rate,
\begin{equation}
\label{hublawsing}
H(t)=f_0\left( t-t_s \right)^{\alpha}\, ,
\end{equation}
where $f_0$ an arbitrary positive constant, $t_s$ an arbitrary time
instance and $\alpha$ a real constant parameter that will be
specified below. We shall assume that $t_s\sim 10^{-35}$, so the
cosmic time $t\sim t_s$ describes an era that belongs to the
early-time evolution. Also note that the parameter $f_0$ may affect
the power spectrum of primordial curvature perturbations. The scale
factor corresponding to the Hubble rate of Eq. (\ref{hublawsing}),
is equal to,
\begin{equation}\label{scalebounce}
a(t)=e^{\frac{f_0}{\alpha+1}\left(t-t_s\right)^{\alpha+1}},
\end{equation}
where we normalized the scale factor, so that $a(t_s)=1$. In order to avoid unnecessary confusion, it is obvious that in order for a bounce to occur, the Hubble rate during the contacting phase should be negative, and also the scale factor should be real for all cosmic times. Hence we assume that $\alpha$ has the following form,
\begin{equation}\label{rev}
\alpha=\frac{2n+1}{2m+1}\,
\end{equation}
where $n$ and $m$ are integers, and hence a bounce occurs and also
the scale factor is real for all cosmic times \footnote{Note that
for example $(-1)^{1/3}$ has two complex branches and one real
negative.}.

The values of the parameter $\alpha$ determine what type of singularity occurs in the evolution. Specifically, the following cases determine the singularity structure of the cosmological evolution:
\begin{itemize}\label{lista}
\item When $\alpha<-1$, the cosmological evolution develops a Type I singularity (Big-Rip) at $t_s$.
\item When $-1<\alpha<0$, the cosmological evolution develops a Type III singularity at $t_s$.
\item When $0<\alpha<1$, the cosmological evolution develops a Type II singularity at $t_s$.
\item When $\alpha>1$, the cosmological evolution develops a Type IV singularity at $t_s$.
\end{itemize}
In the way we chose the Hubble rate (\ref{hublawsing}), the Type IV singularity and the bouncing point both occur at $t=t_s$. Therefore, hereafter when we refer to the bouncing point and to the point that the singularity occurs, we refer to the same thing. In addition we shall assume that $\alpha>1$, in order that the cosmological evolution develops a Type IV singularity.

\section{Description of the Cosmological Evolution and Evolution of Cosmological Scalar Perturbations: The $F(R)$ Gravity Picture}

The singular bounce of Eq. (\ref{hublawsing}), can be generated by a vacuum $F(R)$ gravity (sometimes, called Jordan frame) , as was demonstrated in Ref. \cite{noo5}. Particularly, assuming a vacuum $F(R)$ action of the form,
\begin{equation}
\label{action1dse}
\mathcal{S}=\frac{1}{2\kappa^2}\int \mathrm{d}^4x\sqrt{-g}F(R)\, ,
\end{equation}
and by using the reconstruction techniques of Refs. \cite{Nojiri:2006gh,Capozziello:2006dj,sergbam08}, it was demonstrated in Ref. \cite{noo5}, that the $F(R)$ gravity which describes the singular bounce cosmology of Eq. (\ref{hublawsing}), near the singularity point, that is for $t\to t_s$, is equal to,
\begin{equation}\label{jordanframegravity}
F(R)=R+a_2R^2+a_1 \, ,
\end{equation}
where the parameters $a_1$ and $a_2$ are positive parameters and
their explicit form can be found in the Appendix A. Having at hand
the vacuum $F(R)$ gravity that generates the singular bounce
(\ref{hublawsing}), the focus in this section is twofold. Firstly,
we shall describe in detail the cosmological evolution that the
singular bounce implies, by studying the evolution of the Comoving
Hubble radius $r_H=\frac{1}{a(t)H(t)}$, and secondly we shall
investigate the behavior of the scalar primordial curvature
perturbations near the singularity point, that is near $t\simeq
t_s$. As we demonstrate, the singular bounce picture is completely
different from the inflationary picture.


\subsection{Cosmological Evolution of the Singular Bounce and Comparison with Inflationary Picture}

In order to understand the cosmological evolution in the context of the singular bounce (\ref{hublawsing}), and also the evolution of the corresponding perturbations, it is worth recalling the inflationary picture \cite{inflation,pert}.

The standard Big Bang cosmology was problematic because the problem of initial conditions could not be sufficiently explained, that is, to explain why the Universe appears to be homogeneous in large scale and nearly flat. In addition, there exist large portions of the observed Universe that are similar, with regards to their homogeneity, but these seem that they could not be causally connected in the past, if the Big Bang cosmological evolution is adopted. The inflationary cosmology evolution offered an elegant and consistent description with which, these problems were solved.

Particularly, in the context of inflationary cosmology, the
primordial perturbations of the comoving curvature, which originate
from quantum vacuum fluctuations, which are highly relevant for
present time observations, during the inflationary era were at
subhorizon scales, that is, their wavelength was much smaller than
the Comoving Hubble radius, which is defined to be
$r_H=\frac{1}{a(t)H(t)}$. Since in the context of inflationary
evolution, the Hubble horizon decreases in size during inflation,
these modes freeze once these exit the horizon, with this exit
occurring when the contracting Hubble horizon becomes equal to the
wavelength of these primordial modes. After that, in the classical
inflationary approach, these modes become classical perturbations,
and these superhorizon modes evolve in a classical way. Actually,
these superhorizon modes are conserved after their freezing at the
Hubble horizon exit. Eventually, after the reheating process, and
during the horizon expansion, these initially frozen modes, re-enter
the horizon, and these are actually the modes that are relevant for
present time observations. In effect, the gravitational collapse of
these frozen superhorizon perturbation modes, is responsible for the
formation of our Universe's large-scale structure. Also the
anisotropies of the Cosmic Microwave Background are generated by the
subhorizon modes, which were frozen after the horizon crossing.

As we already mentioned, in the context of the inflationary
evolution,  the Comoving Hubble radius shrinks in a nearly
exponential way. Particularly, at the initial singularity, that
cannot be avoided in the standard inflationary scenario, the
comoving Hubble radius was large and therefore the primordial
perturbations were subhorizon modes, with their comoving wavenumber
satisfying,
\begin{equation}\label{subhorizonsscalaesk}
k\gg H(t)a(t)\, .
\end{equation}
or in terms of the wavelength, $\lambda \ll \left( H(t)a(t)
\right)^{-1}$. Perturbations coming from vacuum fluctuations are
generated during the whole evolution, but the most important for the
present time observations, are the ones with wavelength smaller than
the Comoving Hubble radius. During inflation, the Comoving Hubble
radius decreases dramatically, and at some time instance $t=t_H$,
the perturbations exit the horizon, when $k=a(t_H)H(t_H)$. After the
exit from the horizon, the perturbations freeze and evolve
classically, meaning that the comoving curvature perturbations do
not have a quantum nature anymore, and the quantum expectation value
of the comoving curvature perturbation is the stochastic average of
a classical field. This classical evolution describes perfectly what
``freezing of the modes'' means. After the horizon crossing, the
frozen modes become superhorizon modes, satisfying,
\begin{equation}\label{superhorizon1}
k\ll a(t)H(t)\, .
\end{equation}
After inflation ends, and during the reheating era, the frozen modes reenter the horizon, and this makes possible to have observational data relevant to the inflationary era. Actually, these initially frozen modes correspond to the high energy era, and these are conserved until the horizon reentry, which is the low energy era.

Having described the inflationary picture of cosmological evolution,
let us now describe the cosmological evolution in terms of the
singular bounce appearing in Eq. (\ref{hublawsing}), for which, the
Comoving Hubble radius $r_H(t)=\frac{1}{a(t)H(t)}$ reads,
\begin{equation}\label{hubbleradiusbounce}
r_H(t)=\frac{e^{-\frac{(t-t_s)^{1+\alpha } f_0 }{1+\alpha }} (t-t_s)^{-\alpha }}{f_0 }\, .
\end{equation}
Hence, since for a Type IV singularity, the parameter $\alpha$
satisfies $\alpha>1$, the Comoving Hubble radius $r_H(t)$ strongly
diverges at the bouncing point, which is also the point that the
singularity occurs. Now, this behavior dictates that basically all
the cosmological modes that are important for the present day
observations, were at subhorizon scales at the bouncing point, since
all the modes satisfy $k\gg H(t_s)a(t_s)=0$ at the bouncing point,
or in terms of the wavelength, $\lambda \ll r_H$. For cosmological
times right after the bouncing point, that is $t>t_s$, the comoving
Hubble radius decreases, in an exponential way. For example for two
time instances, $t=10^{-20}$sec and $t=10^{-10}$sec, the fraction of
the Comoving Hubble radius at these two time instances is,
\begin{equation}\label{nearbouncefraction}
\frac{r_H(10^{-20})}{r_H(10^{-10})}\simeq 10^{12}\, .
\end{equation}
The above equation, namely Eq. (\ref{nearbouncefraction}), is very important for our analysis as we shall see. But which modes are the observationally relevant modes? And in addition, does the $F(R)$ gravity description appearing in Eq. (\ref{jordanframegravity}) still describe the near to the bounce cosmological times? This can be easily answered by looking Eq. (\ref{nearbouncefraction}), and by assuming for example that $t_s=10^{-35}$sec, we can clearly see that even modes corresponding to $t=10^{-10}$sec, yield $t-t_s\rightarrow 0$. So practically, the modes which correspond to cosmic times near the bounce, are the observationally relevant for our analysis, since these contain all the relevant information for the early quantum fluctuating vacuum.

The above considerations indicate that the modes for which $t-t_s\rightarrow 0$, practically will determine the spectrum of primordial curvature perturbations, since these are the subhorizon modes at early times. In the following section we shall study in detail the power spectrum of primordial curvature perturbations, and we analyze in detail the observational implications of our analysis.

\begin{figure}[h] \centering
\includegraphics[width=15pc]{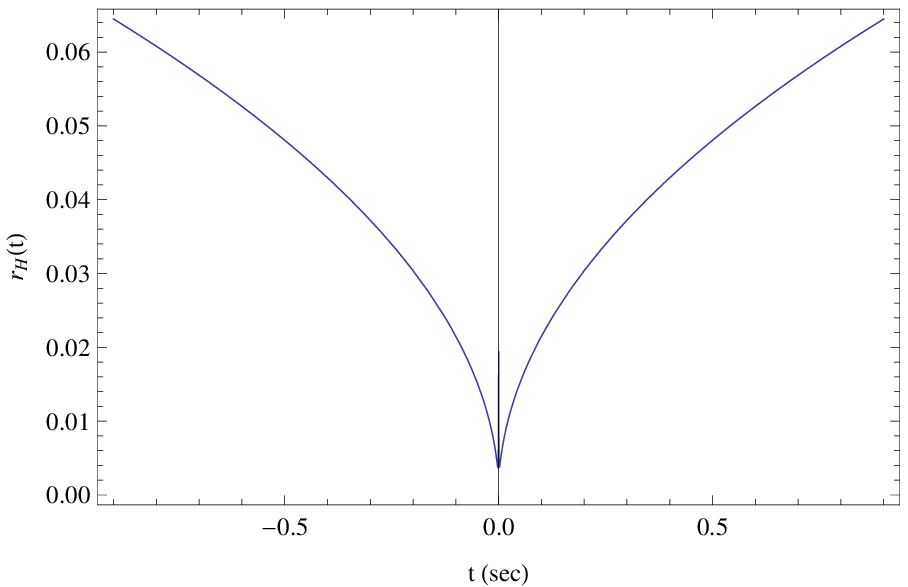}
\includegraphics[width=15pc]{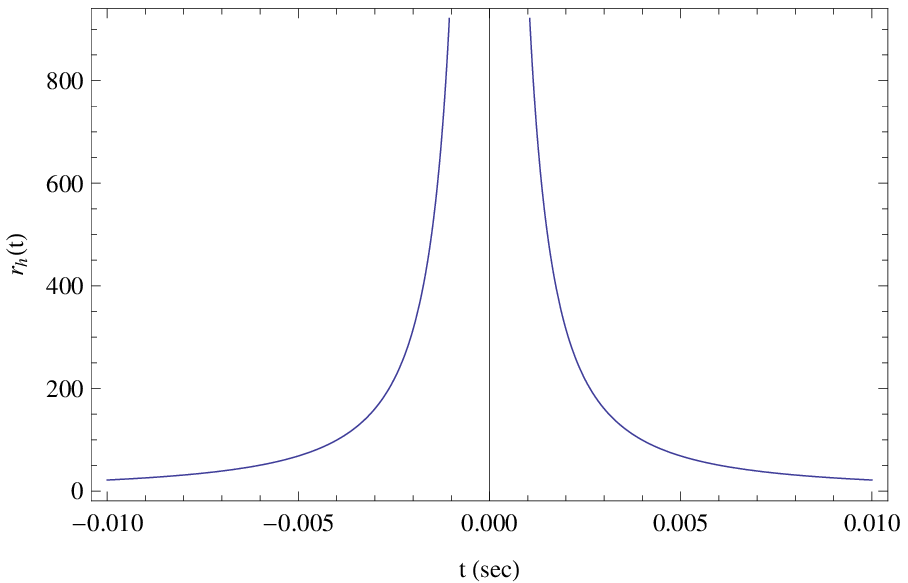}
\caption{The Comoving Hubble radius $r_H(t)$ for the matter bounce
scenario (left plot) and for the singular bounce scenario (right
plot).} \label{singbounce1}
\end{figure}

Before we proceed it is worth discussing an important issue that
will make the analysis that follows more clear. The singular bounce
of Eq. (\ref{scalebounce}) is different in comparison to other
bouncing cosmologies, mainly because the era near the bouncing point
has a peculiar feature. Particularly, the Hubble horizon quantified
in terms of the corresponding Comoving Hubble radius
$r_H(t)=\frac{1}{a(t)H(t)}$, is infinite for cosmic times near the
bouncing point, so we shall assume that the cosmic time is of the
order $t\simeq 10^{-35}$sec, which is very small. This is in
contrast to the Comoving Hubble radius of the matter bounce
scenario, in which case the Comoving Hubble radius is finite near
the bouncing point era. In Fig. \ref{singbounce1} we plotted the
Comoving Hubble radius as a function of the cosmic time, for the
matter bounce scenario (left plot) and for the singular bounce
scenario (right plot).
Therefore the physical interpretation of the era near the bouncing points for the two bouncing cosmologies is totally different.
In the matter bounce scenario, the cosmological fourier modes relevant for present time observations originate at an era much more earlier
 from the bouncing point, and particularly the cosmological perturbations are generated during the contraction era. At that era, for the
 matter bounce scenario, these primordial modes where at subhorizon scales, since during the contraction era, the Hubble horizon shrinks from
 an infinite size, as it can be see from the left plot of Fig. \ref{singbounce1}. In contrast to this physical picture, the singular bounce
 case is totally different. Indeed, at large times, the Hubble horizon is shrunk to zero size, and only for cosmic times near the bouncing point,
 the Hubble horizon has an infinite size. So the primordial modes relevant for present time era are generated for cosmic times near the bouncing point,
 because only at that time all the primordial modes are contained in the horizon. As the horizon shrinks, these modes exit the horizon and these modes
 are relevant for present time observations. Hence the physical picture is very different for the singular bounce, in comparison to other bouncing scenarios.
 We need to note that the behavior of the Hubble radius which we
 discussed above occurs for small times, this is why we assumed that
 the perturbation modes corresponding to $t\ll 10^{-35}$sec will be relevant
 for the calculation of the power spectrum.

This is the reasoning behind our assumption that even modes corresponding to $t\simeq 10^{-10}$sec may belong to the era near the bouncing point. The time instance that the bouncing point may occur is chosen arbitrarily, so we chose $t_s\simeq 10^{-35}$sec, which is the typical time scale for chaotic inflation models at Grand Unified Theories scale, however, this choice is arbitrary. What is important is that the primordial cosmological modes originate at the era which occurs at cosmic times for which $t- t_s\to 0$. Hence, since if we choose $t\simeq 10^{-10}$sec, this choice may also describe primordial modes, since in this case $(10^{-10}-10^{-35})$sec$\simeq 10^{-10}$sec$\ll 1$, so indeed at that time the modes are in subhorizon scales for the singular bounce, since the Hubble horizon is infinite for $t-t_s\to 0$. This is why we claimed that even modes at cosmic times $t\sim 10^{-10}$sec may be relevant for present day observations. Of course the choice is irrelevant, the approximation is only what interests us, the numerical value $t\sim 10^{-10}$sec is just an example.

The fact that the Hubble horizon for cosmic times that satisfy $t-t_s\to 0$, is infinitely large, justifies the approximation that for modes well inside the horizon, a Bunch-Davies vacuum describes these modes. We also discuss this later on.

\subsection{Scalar Perturbations near the Bounce}

In this section we investigate the behavior of the primordial scalar curvature perturbations near the bouncing point, for $t\rightarrow t_s$. The form of the perturbations of the flat FRW background of Eq. (\ref{metricfrw}), is assumed to be of the form \cite{inflation,noh},
\begin{equation}\label{scalarpertrbubations}
\mathrm{d}s^2=-(1+\psi)\mathrm{d}t^2-2a(t)\partial_i\beta\mathrm{d}t\mathrm{d}x^i+a(t)^2\left( \delta_{ij}+2\phi\delta_{ij}+2\partial_i\partial_j\gamma \right)\mathrm{d}x^i\mathrm{d}x^j\, ,
\end{equation}
where the scalar functions $\psi$, $\phi$, $\gamma$ and $\beta$ are smooth scalar perturbations. In principle, perturbations are more easy to analyze in terms of gauge invariant quantities, so for the study of perturbations we shall focus in analyzing the following gauge invariant quantity,
\begin{equation}\label{confedf}
\Phi=\phi-\frac{H\delta \sigma}{\dot{\sigma}}\, ,
\end{equation}
with $\sigma=\frac{\mathrm{d}F}{\mathrm{d}R}$. Note that the gauge invariant quantity $\Phi$ is known as the comoving curvature perturbation. The master equation that describes the evolution of the primordial scalar perturbations of the $F(R)$ gravity theory is,
\begin{equation}\label{perteqnmain}
\frac{1}{a(t)^3Q(t)}\frac{\mathrm{d}}{\mathrm{d}t}\left(a(t)^3Q(t)\dot{\Phi}\right)+\frac{k^2}{a(t)^2}\Phi=0\, ,
\end{equation}
with the term $Q(t)$, for the $F(R)$ gravity case, being equal to \cite{noh},
\begin{equation}\label{gfgdhbhhyhjs}
Q(t)=\frac{\left( \frac{\mathrm{d}^2F}{\mathrm{d}R^2}\right )^2\dot{R}^2}{\kappa^2H(t)^2\left(1+\frac{\frac{\mathrm{d}^2F}{\mathrm{d}R^2}\dot{R}}{2H(t)F(R)}\right)^2}\, .
\end{equation}
where the dot indicates differentiation with respect to the cosmic time $t$. For the singular bounce case of Eq. (\ref{hublawsing}), it is rather difficult to find an analytic solution of differential equation (\ref{perteqnmain}), so we shall seek an approximate solution for cosmological times near the bouncing point, that is for $t-t_s\simeq 0$. Prior proceeding to the approximate solution, after some simple manipulations, the differential equation (\ref{perteqnmain}) becomes,
\begin{equation}\label{difnew}
a(t)^3Q(t)\ddot{\Phi}+\left(3a(t)^2\dot{a}Q(t)+a(t)^3\dot{Q}(t)\right)\dot{\Phi}+Q(t)a(t)k^2\Phi=0\, .
\end{equation}
By using the form of the $F(R)$ gravity appearing in Eq. (\ref{jordanframegravity}), after some tedious algebraic manipulations, the differential equation that describes the evolution of the scalar primordial perturbations near the bouncing point, is the following,
\begin{equation}\label{simplifiedeqn1}
\frac{6}{\kappa ^2}\left(t-t_s\right)\ddot{\Phi}(t)-\frac{12 \alpha }{ \kappa ^2}\dot{\Phi}(t)+k^2 (t-t_s)\Phi (t)=0\, ,
\end{equation}
where the dot denotes differentiation with respect to the cosmic time $t$. The differential equation (\ref{simplifiedeqn1}) has the following analytic solution,
\begin{equation}\label{solutionevolution}
\Phi(t)= \mathcal{C}_1x^{\mu}J_{\mu}(\frac{k x \kappa }{\sqrt{6}})+\mathcal{C}_2x^{\mu} Y_{\mu}(\frac{k x \kappa }{\sqrt{6}})\, ,
\end{equation}
where for convenience we introduced the variable $x=t-t_s$, and in addition $\mu=\frac{1}{2} (1+2 \alpha )$, while $J_{\mu}(t)$ and $Y_{\mu}(t)$ are the Bessel functions of first and second kind respectively. Moreover, $\mathcal{C}_1$ and $\mathcal{C}_2$ are arbitrary integration constants, which be specified later and will be determined by the Bunch-Davies vacuum state for $\Phi$. Since we are interested in the limit $x\rightarrow 0$, we can simplify the expression in Eq. (\ref{solutionevolution}), by taking the limit $x\rightarrow 0$ for the two Bessel functions, which yield,
\begin{equation}\label{besselapprox}
J_{\mu}(y)\simeq \frac{y^{\mu }2^{-\mu }}{\Gamma (1+\mu )},\, \,\,Y_{\mu}(y) \simeq -\frac{2^{-\mu }y^{\mu } \cos(\pi  \mu ) \Gamma (-\mu )}{\pi }-\frac{2^{\mu } \Gamma(\mu )}{\pi }y^{-\mu }\, .
\end{equation}
In the limit $x\rightarrow 0$, the most dominant is the last term, so the evolution of the comoving curvature scalar perturbation near the Type IV singularity becomes approximately equal to,
\begin{equation}\label{approximatebehavior}
\Phi(t)\simeq  -\mathcal{C}_2\frac{2^{3 \mu /2} 3^{\mu /2}\kappa ^{-\mu } \Gamma(\mu )}{\pi }\,x^{\frac{1}{2}+\alpha -2\mu }\,k^{-\mu }   \, .
\end{equation}
Hence in the vacuum $F(R)$ case, the evolution of the comoving curvature perturbation behaves as a power-law function of the variable $x$. The next step is to investigate whether the power spectrum is scale invariant or not, at least near the bouncing point and always having in mind that we only have an approximate expression for both the $F(R)$ gravity and for the comoving curvature perturbation.

The comoving curvature perturbation $\Phi$, which is defined in Eq. (\ref{confedf}), satisfies the following action, which is of second order in perturbation \cite{pert},
\begin{equation}\label{secondoredperturbas}
\mathcal{S}_p=\int \mathrm{d}x^4a(t)^3Q_s\left( \frac{1}{2}\dot{\Phi}-\frac{1}{2}\frac{c_s^2}{a(t)^2}(\nabla \Phi )^2\right)\, ,
\end{equation}
with the variable $Q_s$, being defined in terms of $Q(t)$ as follows, $Q_s=\frac{4}{\kappa^2}Q(t)$ and $Q(t)$ is defined in Eq. (\ref{gfgdhbhhyhjs}). According to the cosmological perturbation theory \cite{inflation,pert}, the power spectrum of the primordial curvature perturbations for the gauge invariant scalar field quantity $\Phi$, is defined as follows,
\begin{equation}\label{powerspecetrumfgr}
\mathcal{P}_R=\frac{4\pi k^3}{(2\pi)^3}\big{|}\Phi\big{|}_{\,\,k=aH}^{\,\,2}\, ,
\end{equation}
which as is obvious from Eq. (\ref{powerspecetrumfgr}), it is evaluated at the horizon crossing, when $k=a(t_H)H(t_H)$, where $t_H$ is the cosmic time at which the horizon crossing occurs. By substituting the comoving curvature perturbation $\Phi$ from Eq. (\ref{approximatebehavior}), the behavior of the power spectrum as a function of the wavenumber $k$, is approximately,
\begin{equation}\label{powerspectrajb}
\mathcal{P}_R\sim k^3 \big{|}\, \mathcal{C}_2\, (t-t_s)^{\frac{1}{2}+\alpha -2\mu }k^{-\mu}\,\big{|}^{\,2}_{\,k=aH}\, .
\end{equation}
At this point it is not possible to determine whether the power spectrum is scale invariant or not, since the condition $k=a(t)H(t)$ and also the constant $\mathcal{C}_2$ contains some $k$-dependence. Therefore, we should further investigate the $k$-dependence of the aforementioned terms. We start off with the term $(t-t_s)^{\frac{1}{2}+\alpha -2\mu }$, which should be evaluated by using the condition $k=a(t)H(t)$. The latter, owing to the fact that for $t-t_s\simeq 0$, yields the following condition,
\begin{equation}\label{explicitequation}
(t-t_s)\simeq k^{\frac{1}{\alpha}}\, ,
\end{equation}
since $a(t)\simeq 1$ and therefore even at the horizon crossing we have $k\simeq H(t)$.

Now we proceed to find the $k$-dependence of the integration constant $\mathcal{C}_2$, and in order to do so we first notice that near the bouncing point, since $t\simeq t_s$, the conformal time $\tau$, which is defined as $\mathrm{d}\tau =a^{-1}(t)\mathrm{d}t$, is identical to the cosmic time, since the scale factor near the bounce satisfies $a(t)\simeq 1$. Hence in the following, where the conformal time is used, we shall assume it is identical to the cosmic time $t$. We shall use this feature of the theory near the bounce, in order to reveal the $k$-dependence of $\mathcal{C}_2$. We introduce the variable $u$ \cite{pert}, which is defined as $u=z_s \Phi$, with $z_s=Q(t)a(t)$, and $Q(t)$ is defined in Eq. (\ref{gfgdhbhhyhjs}). Since near the bounce, $a(t)\simeq 1$, the variable $z_s$ satisfies $z_s\simeq Q(t)$, and hence $u$ satisfies,
\begin{equation}\label{safakebelieve}
u\sim \Phi Q(t)\, .
\end{equation}
Writing the action of Eq. (\ref{secondoredperturbas}) in terms of $u$, it becomes near the bounce,
\begin{equation}\label{actiaonerenearthebounce}
\mathcal{S}_u\simeq \int \mathrm{d}^3\mathrm{d}\tau \left[ \frac{u'}{2}-\frac{1}{2}(\nabla u)^2+\frac{z_s''}{z_s}u^2\right ]\, ,
\end{equation}
with the prime in this case denoting differentiation with respect to the conformal time, which near the bouncing point is identical to the cosmic time, as we evinced earlier. The $k$-dependence of the integration constant $\mathcal{C}_2$ will be determined by the initial vacuum state of the canonical field $u$, which is the Bunch-Davies vacuum state \cite{pert}, at the time instance $t=t_s$, so the field $u$ satisfies,
\begin{equation}\label{bd}
u\sim \frac{e^{-ik\tau}}{\sqrt{k}}\, ,
\end{equation}
where the imaginary phase will be eliminated since the power spectrum is determined by the norm of the comoving curvature $|\Phi (t=t_s)|^2$. The Bunch-Davies vacuum choice is justified since the primordial modes at $t=t_s$ are well inside the Hubble horizon, as it can be seen from Fig. \ref{singbounce1}.

Therefore, by using Eq. (\ref{safakebelieve}), we obtain,
\begin{equation}\label{imanrun}
\Phi(t=t_s)\sim \mathcal{C}_2\sim \frac{1}{\sqrt{k}Q(t)}\, ,
\end{equation}
and since near $t\simeq t_s$, $Q(t)$ is approximately equal to $Q(t)\simeq \frac{6}{\kappa ^2}$, we finally have,
\begin{equation}\label{equationforc4final}
\mathcal{C}_2\simeq \frac{1}{\sqrt{k}}\, .
\end{equation}
By combining Eqs. (\ref{equationforc4final}), (\ref{explicitequation}) and (\ref{powerspectrajb}), the $k$-dependence of the power spectrum $\mathcal{P}_R$, receives the following form,
\begin{equation}\label{powerspectrumfinal}
\mathcal{P}_R\sim k^{3+\frac{1+\alpha-4\mu-2\mu \alpha}{\alpha}}\, .
\end{equation}
Therefore, the power spectrum at a first glance is not scale invariant, but let us investigate if it can be scale invariant for some value of $\alpha$. The spectral index of primordial curvature perturbations $n_s$, is defined in terms of the power spectrum, as follows,
\begin{equation}\label{fieldns}
 n_s-1\equiv\frac{d\ln\mathcal{P}_{\mathcal{R}}}{d\ln
k},
\end{equation}
and hence, for the power spectrum of Eq. (\ref{powerspectrumfinal}), it becomes,
\begin{equation}\label{ns}
n_s=4+\frac{1+\alpha-4\mu-2\mu \alpha}{\alpha}\, .
\end{equation}
The latest Planck data \cite{planck} constraint the spectral index of primordial curvature perturbations as follows,
\begin{equation}\label{recentplancdata}
n_s=0.9655\pm 0.0062\, ,
\end{equation}
and as it can be checked, the spectral index of Eq. (\ref{ns}), cannot be compatible with the value given in Eq. (\ref{recentplancdata}), for no value of $\alpha$.

This result is of some importance, since we concluded that in the case of the singular bounce, if the curvature perturbations relevant for present time observations, originate from the era near the bouncing point, the resulting power spectrum is not scale invariant. This means that the singular bounce is not a cosmologically viable scenario, however we need to be cautious before we conclude this. This is because, the resulting picture is just an approximation near the bounce, which was necessary due to the absence of analyticity.

However, it is possible that the singular bounce itself is not a cosmologically viable scenario. This is an interesting possibility, since the singular bounce might describe the early Universe, and then an alternative cosmological scenario is needed, so that the primordial curvature perturbations are generated during the corresponding era. Such a combined cosmological evolution was described in Ref. \cite{lcdmcai}, see also \cite{lcdmsergei}. In fact, such a scenario is compelling in the singular bounce case, since there are two more reasons indicating that the singular bounce is incomplete. We discuss these issues in the next section.

\subsection{Flaws of the Singular Bounce Evolution and Possible Viable Singular Cosmological Scenarios}

One of the two reasons that the singular bounce might not be an attractive cosmological scenario, unless it is combined with another scenario, is the fact that the primordial curvature perturbations corresponding to the era near the bounce, do not ``freeze'', after the horizon exit. In fact, these are not conserved and actually they grow linearly with time, as we now evince. Before we proceed, we need to note that this feature appears in viable bouncing cosmologies, like for example in the matter bounce scenario \cite{mukhanov1aa}, where the perturbations grow too. After the horizon exit, the wavenumber $k$ satisfies $k\ll a(t)H(t)$, and therefore the last term in the differential equation (\ref{perteqnmain}), that determines the evolution of the perturbations, can be neglected, so it becomes,
\begin{equation}\label{perteqnmainportoriko}
\frac{1}{a(t)^3Q(t)}\frac{\mathrm{d}}{\mathrm{d}t}\left(a(t)^3Q(t)\dot{\Phi}\right)=0\, ,
\end{equation}
and by solving it, the solution becomes,
\begin{equation}\label{soldiffeqn}
\Phi (t)=\mathcal{C}_1+\mathcal{C}_3\int \frac{1}{a(t)^3Q(t)}\mathrm{d}t\, ,
\end{equation}
with the function $Q(t)$ appearing in Eq. (\ref{gfgdhbhhyhjs}), and $\mathcal{C}_1$, $\mathcal{C}_3$ are arbitrary integrations constants. Obviously, the integral in Eq. (\ref{soldiffeqn}) determines whether the perturbations are conserved, and this critically depends on the cosmic times for which the relation $k\ll a(t)H(t)$ holds true. In fact, the cosmic times for which $k\ll a(t)H(t)$, satisfy $t\gg t_s$, but this does not mean that $t\gg 1$. In fact, it is possible that $t-t_s\simeq 0$, although that $t\gg t_s$. For example, if $t\simeq 10^{-15}$sec, it still holds true that $(t-t_s)\simeq 0$, but also that $t\gg t_s$. Hence, since as we evinced earlier, for $t-t_s\simeq 0$, the function $Q(t)$ satisfies $Q(t)\simeq \frac{6}{\kappa^2}$ , and also the scale factor is $a(t)\simeq 1$, we obtain that the integral in Eq. (\ref{soldiffeqn}) yields a linear dependence on $t$,
\begin{equation}\label{soldiffeqnrs}
\Phi (t)=\mathcal{C}_1+\frac{\kappa ^2\mathcal{C}_3}{6}t\, ,
\end{equation}
so the primordial perturbations after the horizon exit, grow
linearly in time. Apart from this rather unwanted feature, there is
also another indication that the singular bounce must be combined
with another cosmological scenario. It is the fact that the Comoving
Hubble radius for the singular bounce always shrinks, and does not
expand. Let us analyze this in some detail.

Every viable cosmological description of our Universe definitely has to describe the inflationary era and after that the reheating process. An important feature of the inflationary era, is that the Hubble horizon shrinks in an exponential way, and after inflation ends, it starts to expand again. In this way, the primordial modes, which freeze at the horizon exit during inflation, reenter the horizon, as this expands at a later time. Note that even today, the era between the horizon exit and reentry, is not fully understood, which means that a complete theory will successfully fill in the theoretical gaps. Unfortunately, one not so appealing feature of the singular bounce, is that after the bounce, the horizon shrinks for all times, so this means that the horizon reentry cannot be described in adequate way. Also, the fact that the power spectrum of the primordial curvature perturbations corresponding to cosmic times near the bounce, is not scale invariant, and cannot be in any physical way scale invariant (unless $\alpha$ is a complex number), this clearly indicates that the singular bounce has to be combined with another cosmological scenario, and together they might yield a viable cosmological description. Such examples have been studied in \cite{lcdmcai} and also \cite{lcdmsergei}. The perturbations in the model of \cite{lcdmcai}, do not correspond to the radiation bouncing phase, but to a much more later time, so scale invariance could be achieved. It might also be possible that the same principle applies for the singular bounce too. In fact, the singular bounce can describe the quantum era, which comes much before the cosmologically relevant for present time observations modes. So in some way, the shrinking horizon might continue until a time instance, much after the bounce, and then the horizon exit might occur. This scenario is quite interesting and will be studied in a future work.

\section{Einstein Frame Picture of the Singular Bounce}

Having discussed the implications of the singular bounce in the Jordan frame, it is worth discussing what is the corresponding picture in the Einstein frame. Particularly it is interesting to see how the singular bounce cosmology with scale factor as in Eq. (\ref{scalebounce}), behaves in the Einstein frame, if it is appropriately conformally transformed. The conformal transformation of the Jordan frame $F(R)$ gravity of Eq. (\ref{jordanframegravity}) results to a canonical scalar field action, with a scalar potential. Particularly, the potential of the $R^2$ Lagrangian of Eq. (\ref{jordanframegravity}) results to a nearly $R^2$ potential, as shown in \cite{sergeistarobinsky}. This kind of potentials were firstly studied in \cite{starobinsky}, and the original Jordan frame $R^2$ theory was firstly studied and introduced in Ref. \cite{starobinsky}. Before discussing how the flat FRW metric of Eq. (\ref{metricfrw}) with scale factor (\ref{scalebounce}) transforms in the Einstein frame, we firstly describe the formalism of the conformal transformation from the Einstein to the Jordan frame. For more details on this see \cite{reviews1}.

The Jordan frame action of the $F(R)$ gravity of Eq. (\ref{jordanframegravity}) is the following,
\begin{equation}
\label{pure} \mathcal{S}=\frac{1}{2\kappa^2}\int
\mathrm{d}^4x\sqrt{-\hat{g}}F(R)=\frac{1}{2\kappa^2}\int
\mathrm{d}^4x\sqrt{-\hat{g}}\left(R+a_2 R^2+a_1 \right)\, ,
\end{equation}
with $\hat{g}_{\mu \nu}$ denoting the Jordan frame metric tensor, the line element of which we assume to be the FRW metric of Eq. (\ref{metricfrw}) with scale factor (\ref{scalebounce}). By introducing the auxiliary scalar field $A$, the Jordan frame action of Eq. (\ref{pure}), can be written in terms of this auxiliary scalar field as follows,
\begin{equation}\label{action1dse111}
\mathcal{S}=\frac{1}{2\kappa^2}\int
\mathrm{d}^4x\sqrt{-\hat{g}}\left ( F'(A)(R-A)+F(A) \right )\, .
\end{equation}
By varying the action (\ref{action1dse111}) with respect to the scalar field $A$, we obtain the solution $A=R$, which implies that the actions of Eqs. (\ref{pure}) and (\ref{action1dse111}), are mathematically equivalent. In order to find the Einstein frame counterpart action of Eq. (\ref{action1dse111}), we perform the following conformal transformation,
\begin{equation}\label{can}
\varphi =-\sqrt{\frac{3}{2\kappa^2}}\ln (F'(A)) \, ,
\end{equation}
In the above equation (\ref{can}), the canonical scalar field $\varphi$ represents the Einstein frame scalar field. Accordingly, the conformal transformation of the Jordan frame metric $\hat{g}_{\mu \nu
}$, is given below,
\begin{equation}\label{conftransmetr}
g_{\mu \nu}=e^{-\varphi }\hat{g}_{\mu \nu } \, ,
\end{equation}
and with $g_{\mu \nu}$ we denote the Einstein frame metric. Under the conformal transformation of Eq. (\ref{conftransmetr}), we can easily obtain the Einstein frame action of the canonical scalar field $\varphi$, which is,
\begin{align}\label{einsteinframeaction}
\mathcal{\tilde{S}}= & \int \mathrm{d}^4x\sqrt{-g}\left (
\frac{R}{2\kappa^2} -\frac{1}{2}\left (\frac{F''(A)}{F'(A)}\right
)^2g^{\mu \nu }\partial_{\mu }A\partial_{\nu }A -\frac{1}{2\kappa^2}
\left ( \frac{A}{F'(A)}-\frac{F(A)}{F'(A)^2}\right ) \right ) \nn =&
\int \mathrm{d}^4x\sqrt{-g}\left ( \frac{R}{2\kappa^2}-\frac{1}{2}g^{\mu
\nu }\partial_{\mu }\varphi\partial_{\nu }\varphi -V(\varphi )\right
)\, ,
\end{align}
where the scalar potential $V(\varphi )$ appearing in Eq. (\ref{einsteinframeaction}), is equal to,
\begin{align}\label{potentialvsigma}
V(\varphi
)=\frac{A}{F'(A)}-\frac{F(A)}{F'(A)^2}=\frac{1}{2\kappa^2}\left (
e^{\sqrt{2\kappa^2/3}\varphi }R\left (e^{-\sqrt{2\kappa^2/3}\varphi}
\right ) - e^{2\sqrt{2\kappa^2/3}\varphi }F\left [ R\left
(e^{-\sqrt{2\kappa^2/3}\varphi} \right ) \right ]\right ) \, .
\end{align}
In the case at hand, for the $F(R)$ gravity of Eq. (\ref{jordanframegravity}), the scalar potential of the canonical scalar field $\varphi$ is equal to,
\begin{equation}\label{vapprox}
V(\varphi)\simeq \frac{1}{4a_2}+\left( \frac{1}{4a_2}-a_1\right)e^{-2\sqrt{\frac{2}{3}}\kappa   \varphi
}-\frac{1}{2a_2}e^{-\sqrt{\frac{2}{3}}\kappa   \varphi } \, .
\end{equation}
This kind of potentials are studied in \cite{sergeistarobinsky}.

Our aim in this section is not to study the conformally transformed Einstein frame scalar theory, but to see how the metric transforms in the Einstein frame, emphasizing on the behavior of the transformed metric near the bouncing point of the Jordan frame metric. Recall that the Jordan frame metric line element is,
\begin{equation}\label{jordanframemetric}
\mathrm{d}s^2=-\mathrm{d}t^2+e^{\frac{2f_0(t-t_s)^{\alpha+1}}{\alpha+1}}\left( \sum_i \mathrm{d}x_i^2\right)\, .
\end{equation}
In order to reveal the behavior of the transformation near the bouncing point, we shall make use of (\ref{can}) and find how this behaves near the bouncing point. Since $F'(R)=1+2a_2 R$ and also $A=R$, the Eq. (\ref{can}), is rewritten as follows,
\begin{equation}\label{cannew}
\varphi =-\sqrt{\frac{3}{2\kappa^2}}\ln (1+2a_2 R)\, .
\end{equation}
Moreover, the Ricci scalar for the singular bounce is,
\begin{equation}\label{ricc}
R=6 \left(2 f_0^2 (t-t_s)^{2 \alpha }+f_0 (t-t_s)^{-1+\alpha } \alpha \right)\, ,
\end{equation}
which for the case of a Type IV singularity it can be approximated as follows,
\begin{equation}\label{ricc1}
R=6 f_0 (t-t_s)^{-1+\alpha } \alpha\, ,
\end{equation}
so the expression in Eq. (\ref{cannew}) can be written in the following way,
\begin{equation}\label{canntransform}
\varphi =-\sqrt{\frac{3}{2\kappa^2}}\ln (1+12 a_2f_0 (t-t_s)^{-1+\alpha } \alpha )\, .
\end{equation}
Under the conformal transformation (\ref{conftransmetr}), by also combining Eq. (\ref{canntransform}), the metric (\ref{jordanframemetric}) becomes,
\begin{equation}\label{einsteinframemetric1}
\mathrm{d}s^2=(1+12 a_2f_0 (t-t_s)^{-1+\alpha } \alpha )\left (-\mathrm{d}t^2+e^{\frac{2f_0(t-t_s)^{\alpha+1}}{\alpha+1}}\left( \sum_i \mathrm{d}x_i^2\right) \right )\, .
\end{equation}
We can rewrite the metric (\ref{einsteinframemetric1}) by using another variable $\tilde{t}$, as follows,
\begin{equation}\label{einsteinftrasnf}
\mathrm{d}\tilde{s}^2=-\mathrm{d}^2\tilde{t}+\tilde{a}(t)^2(\sum_i\mathrm{d}^2x_i)\, ,
\end{equation}
where the variable $\tilde{t}$ is related to $t$ in the following way,
\begin{equation}\label{newvartilde}
\mathrm{d}\tilde{t}=\sqrt{1+12 a_2f_0 (t-t_s)^{-1+\alpha } \alpha}\,\,\,\mathrm{d}t\, ,
\end{equation}
and the scale factor $\tilde{a}(t)$ is related to $a(t)$ as,
\begin{equation}\label{scalaefactorsrelations}
\tilde{a}=\sqrt{1+12 a_2f_0 (t-t_s)^{-1+\alpha } \alpha}\,\,\,a(t)\, .
\end{equation}
By integrating (\ref{newvartilde}), we get,
\begin{equation}\label{approxintegralnew}
\tilde{t}=\int_{t_i}^{t}\sqrt{1+12 a_2f_0 (t-t_s)^{-1+\alpha } \alpha }\,\,\,\mathrm{d}t\, ,
\end{equation}
where $t_i$ some fiducial time. The integral can be approximated near the singularity by the following expression,
\begin{equation}\label{newtildet}
(t-t_s) \simeq \frac{1}{A^{1/(1-\alpha)}}(t-b_s)^{\frac{1}{1-\alpha}}\, ,
\end{equation}
with $b_s$ being equal to $b_s=-A(t_i-t_s)$ and also with $A$ standing for,
\begin{equation}\label{alphadefinition}
A=-\frac{1}{24 \sqrt{3} a_2 f_0 (-3+\alpha ) \alpha  (1+\alpha ) (-5+3 \alpha )}\, .
\end{equation}
By substituting Eq. (\ref{alphadefinition}) in the scale factor relation (\ref{scalaefactorsrelations}), the resulting scale factor $\tilde{a}(\tilde{t})$ is equal to,
\begin{equation}\label{tildescalefactor}
\tilde{a}(\tilde{t})=\sqrt{1+\frac{12a_2f_0 A}{\tilde{t}-b_s}}\,\,\,e^{\frac{f_0}{\alpha+1}A^{\frac{\alpha+1}{1-\alpha}}(\tilde{t}-b_s)^{\frac{\alpha+1}{1-\alpha}}}\, .
\end{equation}
Therefore, the singular bounce metric of Eq. (\ref{jordanframemetric}) which describes a Type IV singular evolution, under a conformal transformation it becomes near the bouncing point,
\begin{equation}\label{tildemetricjordanframemetric}
\mathrm{d}s^2=-\mathrm{d}\tilde{t}^2+\left (1+\frac{12a_2f_0 A}{\tilde{t}-b_s}\right) e^{\frac{2 f_0}{\alpha+1}A^{\frac{\alpha+1}{1-\alpha}}(\tilde{t}-b_s)^{\frac{\alpha+1}{1-\alpha}}}\left( \sum_i \mathrm{d}x_i^2\right)\, .
\end{equation}
The Hubble rate corresponding to the cosmological evolution of Eq. (\ref{tildemetricjordanframemetric}), is equal to,
\begin{equation}\label{Hubbleratefortilde}
H(\tilde{t})=\frac{A f_0 \left(-A^{\frac{2}{-1+\alpha }} (-b_s+\tilde{t})^{\frac{1+\alpha }{-1+\alpha }}+6 a_2 \left((\alpha -1)(\tilde{t}-b_s)^{-1}-2 A^{\frac{1+\alpha }{-1+\alpha }} f_0 (-b_s+\tilde{t})^{\frac{1+\alpha }{-1+\alpha }}\right)\right)}{ (-b_s+12 A a_2 f_0+\tilde{t}) (1-\alpha )}\, ,
\end{equation}
from which is obvious that there exists a finite-time singularity at $\tilde{t}=b_s$, for all values of $\alpha$, with $\alpha >1$.
 Notice that also the scale factor (\ref{tildescalefactor}) also has a finite-time singularity at $\tilde{t}=b_s$,
 so according to the classification of finite-time singularities we presented in section II, the conformally transformed
 Jordan frame singular bounce becomes a FRW metric with a Big Rip finite-time singularity. This is a qualitatively valuable result,
 because for the vacuum $F(R)$ gravity, the Jordan frame Type IV singularity is transformed to a Big Rip in the Einstein frame.
 We need to note that this results holds for cosmic times near $t\simeq t_s$, since only in this case, the $F(R)$ gravity is given
 by Eq. (\ref{jordanframegravity}). Nevertheless the resulting qualitative picture is appealing, since a soft Jordan frame singularity,
 becomes a crushing type singularity in the Einstein frame. Note that, the change of inflation to bounce was also observed in the
 transformation from non-minimal to minimal scalar theory \cite{sasaki1}. On the other hand, it is known that scalar-tensor theory
 with a Big Rip singularity in the Einstein frame, after transformed  to the corresponding $F(R)$ theory,
 it may lead to a complex $F(R)$ gravity \cite{sergeicom}. Also a
 thorough study on the correspondence of finite-time singularities
 was performed in Ref. \cite{annphysod}, where it was shown that in the two
 frames, in principle the finite-time singularities are of different
 type. Also it is possible to see that acceleration in one frame may
 lead to deceleration in the other frame. This correspondence in the
 two frames needs further exploration, since this issue is highly
 non-trivial.

\section{Slow-roll Indices for the $F(R)$ Gravity Description of the Singular Bounce}

The dynamics of the cosmological evolution is determined by the slow-roll parameters \cite{noh}, which indicate whether inflationary evolution occurs and if yes, how much time does it lasts. In the case we study in this paper, there is no inflationary era, since a bounce model is studied, however, the dynamical evolution of the bounce is determined by the slow-roll parameters. In this section, we shall investigate how the slow-roll parameters behave in the context of the singular bounce. We shall calculate these in the context of a modified $F(R)$ gravity, and we shall be interested in the Hubble flow parameters, which were introduced in Ref. \cite{noh}, and these are defined as follows \cite{encyclopedia,noh},
\begin{equation}\label{hubbleflowpar}
\epsilon_1=-\frac{\dot{H}}{H^2},\,\,\,\epsilon_3=\frac{\sigma
'\dot{R}}{2H\sigma },
\,\,\,\epsilon_4=\frac{\sigma''(\dot{R})^2+\sigma ' \ddot{R}}{H\sigma '
\dot{R}}\, ,
\end{equation}
with $\sigma =\frac{\mathrm{d}F}{\mathrm{d}R}$ and this time, the prime denotes differentiation with respect to the Ricci scalar. As we already mentioned, the Hubble flow parameters of Eq. (\ref{hubbleflowpar}) determine the dynamics of the cosmological evolution, and for the case of the inflationary paradigm, if these become of order one, inflation ends. However, for a bouncing cosmology, there is no point to discuss the inflationary dynamics, so we concentrate on the behavior of the Hubble flow parameters, and discuss how these affect the qualitative behavior of the cosmological evolution.

Therefore, we calculate the Hubble flow parameters of Eq. (\ref{hubbleflowpar}), for the case of the singular bounce cosmology of Eq. (\ref{hublawsing}) for the $F(R)$ gravity case, given in Eq.
(\ref{jordanframegravity}). We shall emphasize to the behavior of the Hubble flow parameters near the Type IV singularity. We start off with the parameter $\epsilon_1$, which reads,
\begin{equation}\label{hubserd}
\epsilon_1=-\frac{(t-t_s)^{-1-\alpha } \alpha }{f_0}\, .
\end{equation}
Recall that for a Type IV singularity, the parameter $\alpha $ has to be satisfy $\alpha >1$, so in effect, the parameter $\epsilon_1$ strongly diverges at the Type IV singularity, which occurs at $t=t_s$. The parameter $\epsilon_2$ for an $F(R)$ gravity is zero, so we proceed to the calculation of $\epsilon_3$, which for the singular evolution (\ref{hublawsing}) it reads,
\begin{equation}\label{dgfffdfire}
\epsilon_3=\frac{6 a_1 \alpha  \left(-1+4 f_0 (t-t_s)^{1+\alpha }+\alpha \right)}{(t-t_s) \left(t-t_s+12 a_1 f_0 (t-t_s)^{\alpha } \alpha \right)}
\,
,
\end{equation}
which near the Type IV singularity behaves as follows,
\begin{equation}\label{epsilon3forrsquare}
\epsilon_3\simeq \frac{6 a_1 (-1+\alpha ) \alpha }{(t-t_s)^2}\, .
\end{equation}
Clearly, at the Type IV singularity the parameter $\epsilon_3$ diverges too, as the first Hubble flow parameter $\epsilon_1$, and the same behavior persists for the parameter $\epsilon_4$, which for the singular bounce Hubble rate behaves as,
\begin{equation}\label{epsilon4forrsquare}
\epsilon_4= \frac{(t-t_s)^{-\alpha } \left(f_0 (t-t_s)^{-3+\alpha } (-2+\alpha ) (-1+\alpha ) \alpha +4 f_0^2 (t-t_s)^{-2+2 \alpha } \alpha  (-1+2 \alpha )\right)}{f_0 \left(4 f_0^2 (t-t_s)^{-1+2 \alpha } \alpha +f_0 (t-t_s)^{-2+\alpha } (-1+\alpha ) \alpha \right)}\, ,
\end{equation}
which can be simplified near the Type IV singularity as follows,
\begin{equation}\label{neartypeivepsilon4}
\epsilon_4\simeq \frac{(t-t_s)^{-1-\alpha } (-2+\alpha )}{f_0}\, ,
\end{equation}
from which we can see that for $\alpha>1$, the Hubble flow parameter $\epsilon_4$ diverges too at $t=t_s$. The resulting physical picture is the following: for the singular bounce Hubble rate (\ref{hublawsing}), the Hubble flow parameters (\ref{hubbleflowpar}) strongly diverge at the Type IV singularity point $t=t_s$. This behavior is compelling and should be interpreted correctly. The singularities at the Hubble flow parameters indicate that the dynamics of the cosmological evolution is abruptly interrupted at the Type IV singularity. This means that at the Type IV singularity, the cosmological solution that described the cosmological evolution up to that point, ceases to be the final attractor of the theory, and the cosmological dynamical system will choose another solution to be it's final attractor. Of course in the context of the singular bounce, we cannot see the complete theory that could describe this picture, but a complete theoretical description, to which the singular bounce would be a part, could effectively describe the aforementioned behavior. The fact that the theory of the singular bounce ceases to be the final attractor of the theory, is also supported by the fact that the resulting power spectrum originating from the era near the bounce, is not and cannot be scale invariant. Hence, as these two results indicate, if the singular bounce is to be considered a viable part of a cosmological evolution, it has to be combined with another cosmological scenario yet to be found, in the context of which, the power spectrum will be scale invariant and also this new cosmological solution will be the final attractor of the theory.

We need to clarify one issue before we continue. We mentioned in the text above that the Type IV singularity affects in a crucial way the dynamical evolution of the cosmological model under study. However, this can be in contrast to the claim that the Universe may pass smoothly through a Type IV singularity, therefore we need to further clarify this seemingly oxymoron claims. When we say that the Universe passes smoothly through a Type IV singularity, we mean that the observable physical quantities classified earlier, which are, the scale factor, the Hubble rate, the effective pressure, the effective energy density and the curvature scalars, are finite at a Type IV finite time singularity. However, the quantities that quantify the dynamical evolution may develop instabilities, so the dynamics of the evolution is affect by the Type IV singularity and this occurs only in some cases and not always. Therefore with the terminology smooth passage, we mean that the physical quantities we just mentioned are finite.

\subsection{The Hubble Slow-Roll Parameters for a Type IV Singular
Evolution}

Let us now calculate another set of slow-roll parameters which are quite frequently used in the literature, which are the Hubble slow-roll parameters \cite{barrowslowroll}. We consider these in the context of the Jordan frame $F(R)$ gravity we studied in this paper, appearing in Eq. (\ref{jordanframegravity}), and we are mainly concerned in finding their behavior near the Type IV singularity. As in the Hubble flow parameters case we studied in the previous section, the Hubble slow-roll parameters also determine the dynamical evolution of the cosmological system under study. As we shall demonstrate, the Hubble slow-roll parameters also become singular at the Type IV singularity, as was probably expected. The Hubble slow-roll parameters are two parameters, which we denote as $\epsilon_H$ and $\eta_H$, and these are equal to,
\cite{barrowslowroll},
\begin{equation}\label{hubbleslowrooll}
\epsilon_H=-\frac{\dot{H}}{H^2},\,\,\, \eta_H=-\frac{\ddot{H}}{2H\dot{H}}\, .
\end{equation}
So by using the Hubble rate of Eq. (\ref{hublawsing}), the first Hubble slow-roll parameter $\epsilon_H$ becomes,
\begin{equation}\label{epsilonforsingularevolution}
\epsilon_H=-\frac{(t-t_s)^{-1-\alpha } \alpha }{f_0}\, ,
\end{equation}
and correspondingly, the slow-roll parameter $\eta_H$ becomes equal to,
\begin{equation}\label{etasingualarhubla}
\eta_H=-\frac{(t-t_s)^{-1-\alpha } (-1+\alpha )}{2 f_0}\, .
\end{equation}
From both equations (\ref{epsilonforsingularevolution}) and (\ref{etasingualarhubla}), it is obvious that in the case of a Type IV singularity at $t=t_s$, the Hubble slow-roll parameters $\epsilon_H$ and $\eta_H$ diverge at the singularity point, and hence a strong instability occurs in the dynamical evolution of the cosmological system. Hence, as in the case of the Hubble flow parameters, the divergent Hubble slow-roll parameters indicate dynamical instability.

Thus, in contrast to the other types of singularities, the Type IV singularity does not result to infinite physical quantities at the singularity point, but the effect of the Type IV singularity appears at the level of the dynamical evolution, which becomes abruptly interrupted at the singularity point.

Before we close this section, and important remark is on order. In principle, someone could claim that the singularities in the Hubble slow-roll parameters would lead to a singular spectral index of primordial curvature perturbations $n_s$, via the equation,
\begin{equation}\label{egefda}
n_s\simeq 1-4\epsilon_H +2\eta_H\, .
\end{equation}
However, this is not true, since the above equation (\ref{egefda}) is valid only when the Hubble slow-roll parameters satisfy $\epsilon_H\ll 1$ and $\eta_H\ll 1$, that is, when the slow-roll condition holds true. Obviously if the slow-roll condition does not hold true for the Hubble slow-roll parameters, then the only way of consistently calculating the spectral index of primordial curvature perturbations $n_s$, is by calculating the power spectrum of the primordial curvature perturbations, as we did in a previous section, and from it, by using Eq. (\ref{fieldns}), the spectral index can be easily and consistently calculated. As it can be seen from Eq. (\ref{ns}), the Type IV singularity has no effect on the spectral index $n_s$, if the latter is correctly calculated. Of course, the relation (\ref{egefda}) can be safely used for time instances away from the singularity, in which case the result for $n_s$ is consistent, but the appearance of singularities should not be misinterpreted.

\section{Singular Evolution Unifying Early-time and Late-time: The Modified Starobinsky Model Case}

As we discussed in the previous sections, the singular bounce cosmology with Hubble rate (\ref{hublawsing}), cannot describe a viable a singular cosmology by itself, since the quantitative features of the theory, with regards to observational quantities, are not appealing at all. Hence, it has to be combined with other cosmological scenarios, for example by being a part of a viable cosmological evolution, describing perhaps the quantum era. But in general, the singular evolution can be generalized in such a way, so that viable cosmological evolution is generated, for which compatibility with Planck data is achieved. In addition, in such a successful singular description, the qualitative features of a viable cosmological evolution, such as the early and late-time acceleration, and also the matter domination and radiation domination eras, have to be described successfully. In this section we briefly present a generalization of the singular evolution (\ref{hublawsing}), to support our argument that in general, the Type IV singular cosmology can have quite appealing properties.

Consider the following cosmological evolution,
\begin{equation}\label{singgener}
H(t)=\frac{2}{3 \left(\frac{4}{3 H_0}+t\right)}+\e^{-(t-t_s)^{\gamma }}
\left(\frac{H_0}{2}+H_i (t-t_i)\right)+f_0 (t-t_0)^{\delta } (t-t_s)^{\gamma
}\, ,
\end{equation}
with $t_s$, $H_0$, $t_0$, $\gamma$, $\delta $, $H_i$, $f_0$ and $t_i$, constant parameters that will be chosen appropriately in order for the Hubble rate (\ref{singgener}) to describe a viable cosmology. The Hubble rate of Eq. (\ref{singgener}) has the same singularity structure as the Hubble rate of the singular bounce, therefore, for $\delta >1$ and $\gamma>1 $, the cosmological evolution develops two different Type IV singularities, at  $t=t_s$ and at $t=t_0$ respectively. The cosmological evolution is chosen in such a way so that it can describe early-time and late-time acceleration and also the matter domination era. In order to see this, we assume that the time instance $t_s$ occurs at the end of the inflationary era, and also the time instance occurs at a much more later time than $t_s$, that is $t_0\gg t_s$. Particularly, the time instance $t_0$ is assumed to occur during the intermediate era between the present time and the end of the inflationary era, and therefore $t_0\ll t_p$. The parameters $H_0$ and $H_i$ are assumed to satisfy $H_0,H_i\gg 1$, and particularly, these are chosen to be identical to the parameters appearing in the Jordan frame $R^2$ inflation model, see for example Ref. \cite{noo4}. Let us proceed to investigate what kind of cosmology does the model of Eq. (\ref{singgener}) describes, starting with early times. As we assumed, the first Type IV singularity occurs at $t=t_s$, which is assumed to occur at early times, so for $t\sim t_s$, the Hubble rate (\ref{singgener}) becomes,
\begin{equation}\label{approx}
H(t)\simeq H_0+H_i(t-t_i)\, ,
\end{equation}
owing to the fact that the term $\sim \frac{2}{3 \left(\frac{4}{3 H_0}+t\right)}$ becomes approximately equal to $\sim \frac{H_0}{2}$, and in addition the term that contains the two Type IV singularities, namely $\sim
(t-t_s)^{\gamma}(t-t_0)^{\delta }$, vanishes at $t\simeq t_s$. Hence at early-time, the singular cosmology of Eq. (\ref{singgener}), becomes approximately the Starobinsky $R^2$ inflation model \cite{starobinsky,noo4}, which is compatible to the observational data coming from Planck \cite{planck}. This feature is very appealing, since the two singularities were eliminated at early-time, leaving only the $R^2$ inflation model, thus the singular cosmological evolution (\ref{singgener}) can potentially generate a nearly scale invariant power spectrum.

Proceeding to the study of later times, the Hubble rate near $t\simeq t_0$, is approximately equal to,
\begin{equation}\label{matterera}
H(t)\simeq \frac{2}{3 \left(\frac{4}{3 H_0}+t\right)}\, ,
\end{equation}
since the term $\sim \e^{-(t-t_s)^{\gamma }}
\left(\frac{H_0}{2}+H_i (t-t_i)\right)$ becomes exponentially suppressed, while the term $\sim (t-t_0)^{\delta }$, is approximately equal to zero. The Hubble rate describes the matter domination era, as it can be easily seen if the effective equation of state (EoS) of the cosmological system is evaluated. So in order to describe the phenomenological picture better, we calculate the EoS for the Hubble rate (\ref{singgener}), which reads,
\begin{align}
\label{generaleosfor modstarmod}
w_{\mathrm{eff}}=&-1-\frac{2 \left(\e^{-(t-t_s)^{\gamma }} H_i-\frac{2}{3
\left(\frac{4}{3 H_0}+t\right)^2}+f_0 (t-t_0)^{\delta } (t-t_s)^{-1+\gamma }
\gamma \right)}{3 \left(\frac{2}{3 \left(\frac{4}{3
H_0}+t\right)}+\e^{-(t-t_s)^{\gamma }} \left(\frac{H_0}{2}+H_i
(t-t_i)\right)+f_0 (t-t_0)^{\delta } (t-t_s)^{\gamma }\right)^2} \nn
& -\frac{2 \left(-\e^{-(t-t_s)^{\gamma }} \left(\frac{H_0}{2}+H_i (t-t_i)\right)
(t-t_s)^{-1+\gamma } \gamma +f_0 (t-t_0)^{-1+\delta } (t-t_s)^{\gamma } \delta
\right)}{3 \left(\frac{2}{3 \left(\frac{4}{3 H_0}+t\right)}+\e^{-(t-t_s)^{\gamma
}} \left(\frac{H_0}{2}+H_i (t-t_i)\right)+f_0 (t-t_0)^{\delta } (t-t_s)^{\gamma
}\right)^2}
\, .
\end{align}
It can be easily checked that at early times, which correspond to cosmic times $t\simeq t_s$, the EoS has the following form,
\begin{equation}\label{eosearlyts}
w_{\mathrm{eff}}\simeq -1-\frac{2 \left(\frac{3 H_0}{4}+H_i\right)}{3 (H_0+H_i
(t-t_i))^2}\, .
\end{equation}
As we discussed earlier, the parameters $H_0$ and $H_i$ are chosen to satisfy $H_0, H_i\gg 1$, so the EoS of Eq. (\ref{eosearlyts}) becomes $w_{\mathrm{eff}}\simeq -1$ and hence, as we discussed earlier too, the early-time era is described by a de Sitter acceleration. At cosmic times of the order $t\simeq t_0$, which occurs much more later than the time $t_s$ and much more earlier than the present time $t_p$, the EoS takes the following form,
\begin{equation}\label{eosearlyts1}
w_{\mathrm{eff}}\simeq -1-\frac{2 \left(\e^{-(t-t_s)^{\gamma }} H_i-\frac{2}{3
\left(\frac{4}{3 H_0}+t\right)^2}-\e^{-(t-t_s)^{\gamma }}
\left(\frac{H_0}{2}+H_i (t-t_i)\right) (t-t_s)^{-1+\gamma } \gamma \right)}{3
\left(\frac{2}{3 \left(\frac{4}{3 H_0}+t\right)}+\e^{-(t-t_s)^{\gamma }}
\left(\frac{H_0}{2}+H_i (t-t_i)\right)\right)^2}\, ,
\end{equation}
and owing to the fact that $t\sim t_0\gg t_s$, and in addition $t\gg \frac{4}{3 H_0}$, Eq. (\ref{eosearlyts1}) becomes,
\begin{equation}\label{eosearlyts1a}
w_{\mathrm{eff}}\simeq -1-\frac{2 \left(\e^{-t^{\gamma }} H_i-\frac{2}{3
t^2}-\e^{-t^{\gamma }} \left(\frac{H_0}{2}+H_i (t-t_i)\right) t^{-1+\gamma }
\gamma \right)}{3 \left(\frac{2}{3 t}+\e^{-t^{\gamma }} \left(\frac{H_0}{2}+H_i
(t-t_i)\right)\right)^2}\, .
\end{equation}
Owing to the fact that for $t\gg 1$, the terms containing the exponential $\sim \e^{-t^{\gamma }}$, become exponentially suppressed, the EoS becomes,
\begin{equation}\label{eosearlyts2}
w_{\mathrm{eff}}\simeq -1-\frac{2 \left(-\frac{2}{3 t^2}\right)}{3
\left(\frac{2}{3 t}\right)^2}= 0\, ,
\end{equation}
and hence the cosmological evolution is described by a nearly a matter dominated era, since $w_{\mathrm{eff}}\simeq 0$. Correspondingly, at late times, the EoS can be approximated as follows,
\begin{equation}\label{eosearlyts3}
w_{\mathrm{eff}}\simeq -1-\frac{2 \left(\e^{-t^{\gamma }} H_i-\frac{2}{3
t^2}+f_0 t^{-1+\gamma +\delta } \gamma - \e^{-t^{\gamma }} t^{-1+\gamma }
\left(\frac{H_0}{2}+H_i (t-t_i)\right) \gamma +f_0 t^{-1+\gamma +\delta }
\delta \right)}{3 \left(\frac{2}{3 t}+f_0 t^{\gamma +\delta }+ \e^{-t^{\gamma }}
\left(\frac{H_0}{2}+H_i (t-t_i)\right)\right)^2}\, ,
\end{equation}
and by omitting the terms which are exponentially suppressed, but also subdominant terms too, the EoS takes the following form,
\begin{equation}\label{eosearlyts4}
w_{\mathrm{eff}}\simeq -1-\frac{2 t^{-1-\gamma -\delta } \gamma }{3
f_0}-\frac{2 t^{-1-\gamma -\delta } \delta }{3 f_0}\, .
\end{equation}
Owing to the fact that $t\gg 1$, the terms containing $\sim t^{-1-\gamma -\delta }$, are subdominant, since $\gamma,\delta>1$ in order for Type IV singularities to occur. Thereby, at late times, the EoS is approximately equal to, $w_{\mathrm{eff}}\simeq -1$, which means that the late-time era s also an accelerating era. Therefore, the combination of two singular evolutions can lead to a quite appealing cosmological evolution with the following features:
\begin{itemize}
    \item Unified description of early and late-time acceleration.
    \item Successful description of the matter domination era.
    \item Compatibility with observational data.
\end{itemize}
However, the brief presentation of the model we performed in this
section deserves a thorough analysis, especially the calculation of
the power spectrum, the evolution of the Comoving Hubble radius
(horizon) and also the exact generation of the model from a modified
$F(R)$ gravity. The full analysis of a model similar to the one we
discussed in this section is performed in Ref. \cite{newref}.

\section{Conclusions}

In this paper we studied a bounce cosmology with a Type IV
singularity occurring at the bouncing point. Our study involved a
simple bounce model, and we studied it in the context of modified
$F(R)$ gravity theory. Particularly, we focused our study on cosmic
times near the Type IV singularity, and we found the approximate
form of the $F(R)$ gravity that describes the bounce near the Type
IV singularity. The Type IV singularity is not a crushing type
singularity, like the Big Rip or the initial singularity, so all the
physical quantities that can be defined on the three dimensional
spacelike hypersurface defined at $t=t_s$, are finite and therefore
the effects of the singularity cannot be seen at the level of
physical observable quantities. Moreover, we analyzed the evolution
of primordial curvature perturbations, in the context of the
approximate $F(R)$ gravity that generates the bounce. Particularly,
after properly defining the comoving curvature perturbation, we
found the quantum action that this gauge invariant quantity
satisfies, and also we presented the differential equation that
governs it's evolution. After investigating the evolution of the
Comoving Hubble radius, or horizon, for the singular bounce, we
determined which modes are relevant for present time observations.
We concluded that for this singular bounce scenario, the relevant
modes are those which correspond at early times, near the bouncing
point. The Hubble horizon for the singular bounce has an infinite
radius at the singularity point and shrinks immediately after the
singularity point, thus at some instance the relevant for current
time observations modes, exit the horizon. The calculation of the
power spectrum and of the corresponding spectral index of primordial
curvature perturbations, revealed that the power spectrum is not
scale invariant and cannot be for any choice of the free parameters
of the theory. Therefore, the singular bounce model in the context
of $F(R)$ modified gravity is problematic to some extent. In
addition, there are also three more reasons for which the singular
bounce model is problematic, which are the following:
\begin{itemize}
    \item The Hubble horizon after the exit of the short wavelength modes, continuously shrinks, without the possibility that it stops shrinking at some point. This is a problem, since in the standard inflationary theory, after the primordial modes exit the Hubble horizon, these freeze, and the horizon after some time, starts to expand again, after the reheating. Hence, these modes reenter the horizon, and this is why these primordial modes become cosmologically relevant for today's observations. Hence, the problem of a continuously shrinking horizon is one of the reasons that the singular bouncing cosmology is problematic.
    \item The second reason that renders the singular bounce not so appealing, is the fact that the resulting power spectrum, generated at cosmic times near the Type IV singularity, is not scale invariant.

\item The third reason is that the short wavelength primordial modes, after these exit the horizon, they do not freeze, but grow in time, in a linear way.

\item The predictability of the theory is limited since the
perturbations grow after the horizon crossing, and these are not
conserved. Hence by calculating simply the power spectrum at the
horizon crossing does not suffice. Regardless of that however, the
model does not produce a scale invariant spectrum in its $F(R)$
gravity, so it is not viable anyway.
\end{itemize}
So the next step for the singular bounce is that, in order to be considered viable, it has to be combined with another cosmological scenario, which will eventually generate the primordial perturbations which are scale invariant, and the short wavelength modes will freeze after the horizon exit. In addition, the horizon will eventually expand again, and therefore the frozen modes will reenter the horizon again. Such a scenario in the context of the radiation bounce scenario of Loop Quantum Cosmology \cite{LQC}, was presented in Ref. \cite{lcdmcai}. In some sense, it seems that the singular bounce will describe the quantum era of the cosmological system, and this era will be followed by another cosmological era, which will generate a scale invariant power spectrum and will be responsible for all the rest features, which the singular bounce itself fails to describe.

We need to note however, that our study involved an approximate form of the $F(R)$ gravity, and therefore, we should investigate if any modification on the form of the $F(R)$ gravity, will change the resulting picture. For example, if the Loop Quantum Cosmology corrected $F(R)$ gravity \cite{sergeifrbounce} is used, it may be possible that the spectrum might be scale invariant, in which case, the results of this study are just an artifact of the approximations we made for the $F(R)$ gravity near the bouncing point. In addition, perhaps the linear perturbation theory analysis breaks down even in the case of the Type IV singularity, or it is just that the power spectrum is not scale invariant.

Another important feature of the Type IV singularity is that it does not result to any singular observable quantity, but affects crucially the dynamical evolution of the cosmological system. Particularly, we calculated the slow-roll indices, which determine the dynamical evolution, and as we demonstrated, these become singular at the singularity point. This clearly indicates that cosmological solution which governed the cosmological system, ceases to be the final attractor of the theory, and therefore the singular bounce, which may described the system for $t<0$ and until $t=t_s$, ceases to describe the system for $t>t_s$, since it becomes dynamically unstable. This result supports our claim that the singular bounce has to be combined with another scenario in order  to be viable, since the singular bounce itself is unstable.

An interesting feature of the Jordan frame singular bounce, is that when the theory is conformally transformed in the Einstein frame, the corresponding Einstein frame metric develops a Big Rip singularity. So for times near the bounce, the Jordan frame Type IV singular bounce becomes a Big Rip singularity in the Einstein frame. This feature certainly deserves some attention, and a study on this issue will be presented in a future work.

As a final task, we briefly investigated a generalized cosmological model, which contained two Type IV singularities. As we demonstrated, the resulting picture was quite appealing, since we were able to describe with the same model, early-time and late-time acceleration, but also the matter domination era. In order to see this, we calculated the effective equation of state of the cosmological system, and the behavior we discussed indeed holds true. The cosmological model itself is a generalization of the singular bounce model, and the singular part becomes relevant only at late times. Also the observational data can be compatible with this model, since the model is similar to the Starobinsky $R^2$ inflation model, which is compatible to the Planck data. However, the study was brief and deserves a more thorough analysis, which we intend to do in the future.

At this point it is worth highlighting the most important outcomes
we found in this work, in order to compare the present work with
previous works:

\begin{itemize}
  \item We examined a soft type of singularity and we
  investigated the phenomenological implications on a non-crushing
  type singular bounce. In most works appearing in the literature,
  the bounces are not singular, but in our case a soft type
  singularity, namely a Type IV singularity, occurs at the bouncing
  point. This is the first time in the literature that a soft type singularity occurs
  for a bounce scenario.
  \item We investigated how the singularity affects the power
  spectrum, and most importantly we addressed in detail the question
  referring in which era the primordial perturbations are actually
  generated. In contrast to the existing bouncing cosmology
  scenarios, in the singular bounce case, the perturbations are
  generated during the beginning of the expanding phase, and not
  during the contracting phase.
  \item We examined the viability of the singular bounce in the
  context of $F(R)$ gravity, and in contrast to the same problem in
  the context of $F(G)$ gravity, which was performed in \cite{noo3}, in
  the case at hand the primordial power spectrum is not scale
  invariant, and the spectral index is not compatible with the
  Planck data.
  \item We also calculated the evolution of perturbations after the
  horizon crossing, and as we showed, the perturbations grow, a fact
  that renders the present singular Type IV singular bounce
  scenario, invalid, at least in the context of F(R) gravity, in
  contrast to the $F(G)$ case \cite{noo3}.

  \item Finally, we argued that the singular bounce scenario should
  be combined with some sort of post-bounce inflationary scenario,
  in order for the primordial power spectrum to be compatible with
  the data. These kind of scenarios often appear in the literature \cite{Liu:2013kea,Piao:2003zm}, so it might be possible that this bounce-inflation combination might yields interesting
  results.

\end{itemize}

In conclusion, the singular bounce model may be a part of a viable cosmological evolution, but cannot be itself a correct description of our Universe for all times. The interesting task is to find what form should have the cosmological scenario that will follow after the singular bounce, with the most possible one is that of a variant $\Lambda$ Cold Dark Matter model. We hope to thoroughly address these issues in a future work.

\section*{Acknowledgments}

This work is supported by MINECO (Spain), project
 FIS2013-44881 (S.D.O) and by Min. of Education and Science of Russia (S.D.O
and V.K.O).

\section*{Appendix: The Explicit Forms of the Parameters $a_1$ and $a_2$ Appearing in Text}

Here we quote the explicit form of the parameters $a_1$ and $a_2$ appearing in Eq. (\ref{jordanframegravity}). Particularly, these are given in terms of the parameters $\mathcal{C}$, $\mathcal{A}$ and $\mathcal{B}$, as follows,
\begin{equation}\label{coeffcnew}
a_2=4\left(-\frac{\mathcal{C}}{4\mathcal{A}^2}\right)^{-1},{\,}{\,}{\,}a_1 =-\frac{\mathcal{B}^2}{\mathcal{C}}+\mathcal{C}\, .
\end{equation}
In turn, the parameters $\mathcal{C}$, $\mathcal{A}$ and $\mathcal{B}$ as proved in \cite{noo5}, are equal to,
\begin{equation}
\label{parametalpha} \mathcal{A}=\frac{f_0^{-\frac{1}{2 (1+\alpha )}}(2+2 \alpha )^{\frac{1}{2 (1+\alpha )}} C_1\Gamma\left(\frac{1}{2 (1+\alpha )}\right)}{2 (1+\alpha ) \Gamma\left(-\frac{1+4 \alpha }{2 (1+\alpha )}\right)}+\frac{f_0^{-\frac{1}{2 (1+\alpha )}} C_1(2+2 \alpha )^{\frac{1}{2 (1+\alpha )}} \left(1+5 \alpha +2 \alpha ^2\right) \Gamma\left(1+\frac{1}{2 (1+\alpha )}\right)}{2 (1+\alpha )^2 \Gamma\left(1-\frac{1+4 \alpha }{2 (1+\alpha )}\right)}\, ,
\end{equation}
with regards to $\mathcal{A}$, while the rest of the parameters, namely $\mathcal{B}$ and $\mathcal{C}$, are equal to,
\begin{align}
\label{mathbandc} \mathcal{B}=&6 f_0 (1+4\alpha ) C_1
\Big{(}\frac{f_0}{2(\alpha+1) }\Big{)}
{}^{-\frac{1}{2(\alpha+1) }} \\ \notag & \times
\frac{\Gamma\Big{(}1+\frac{1}{2(\alpha+1) }\Big{)}
\Big{(}(2+2(\alpha+1) ) -2 (1+ 2(\alpha+1) )
\Gamma\Big{(}\frac{1-2\alpha }{2(\alpha+1)
}\Big{)} \Big{)} }{ (2(\alpha+1) )^2 \Gamma\Big{(}\frac{1-2\alpha }{2(\alpha+1) }\Big{)}} \, , \nn \mathcal{C}=& \frac{6
f_0^{\frac{22(\alpha+1)+1}{2(\alpha+1)+1}} C_1
}{(2(\alpha+1) )^{3+2(\alpha+1)}} \\ \notag & \times
\Big{(}\frac{(2(\alpha+1) ) (2+2(\alpha+1) )
 \Gamma\Big{(}\frac{1}{2(\alpha+1) }\Big{)}}{\Gamma\Big{(}\frac{1+4\alpha }{2(\alpha+1) }\Big{)}}-\frac{2(\alpha+1)  (1+4\alpha )^2
  \Gamma\Big{(}1+\frac{1}{2(\alpha+1) }\Big{)}}{\Gamma\Big{(}\frac{1-2\alpha }{2(\alpha+1) }\Big{)}}
  \\ \notag & -\frac{4 \Big{(}1+ 2(\alpha+1) +2(\alpha+1) ^2\Big{)}
   \Gamma\Big{(}1+\frac{1}{2(\alpha+1) }\Big{)}}{\Gamma\Big{(}\frac{1-2\alpha }{2(\alpha+1) }\Big{)}}+\frac{4 (1+2(\alpha+1) )^2 (1+4\alpha )
   \Gamma\Big{(}2+\frac{1}{2(\alpha+1) }\Big{)}}{(1+2(\alpha+1) )
    }\Big{)} \, .
\end{align}
Note that $C_1$ is an arbitrary parameter, which is equal to \cite{noo5},
\begin{align}\label{c1valueforeinsteihilber}
& C_1=-\frac{(1+2(\alpha+1) )^2 \Gamma\Big{(}\frac{2+4
2(\alpha+1) }{1+2(\alpha+1) }\Big{)}
 \Gamma\Big{(}\frac{3+5 2(\alpha+1) }{1+2(\alpha+1) }\Big{)}}{12 f_0 (1+3 2(\alpha+1) )
 }
 \\ \notag & \times  \frac{1}{\Big{(}(2+2(\alpha+1) ) \Gamma\Big{(}\frac{3+5 2(\alpha+1) }{1+2(\alpha+1) }\Big{)}
 \Gamma\Big{(}1+\frac{1}{1+2(\alpha+1) }\Big{)}-2 (1+2 2(\alpha+1) ) \Gamma\Big{(}\frac{2+4 2(\alpha+1) }{1+2(\alpha+1) }\Big{)} \Gamma\Big{(}2+\frac{1}{1+2(\alpha+1) }\Big{)}\Big{)}
 \Big{(}\frac{f_0}{1+2(\alpha+1) }\Big{)}{}^{-\frac{1}{1+2(\alpha+1) }}}\, ,
\end{align}
which is imposed by the constraint,
\begin{equation}
\label{cond1-B}
-2\frac{\mathcal{B}\mathcal{A}}{\mathcal{C}}=1\, .
\end{equation}

\end{document}